\documentclass[letterpaper,twocolumn,fleqn]{article} 

\usepackage{ist}
\usepackage{amsmath}
\usepackage{multirow}
\usepackage{appendix}
\usepackage{pdfpages}

\title{Defining Cost Function of Steganography with Large Language Models}

\author{Hanzhou Wu and Yige Wang\\
School of Communication and Information Engineering, Shanghai University, Shanghai 200444, China\\
Emails: h.wu.phd@ieee.org, 24721152@shu.edu.cn}

\date{} 

\begin{document} 

\maketitle 

\thispagestyle{empty} 


\begin{abstract}
In this paper, we make the first attempt towards defining cost function of steganography with large language models (LLMs), which is totally different from previous works that rely heavily on expert knowledge or require large-scale datasets for cost learning. To achieve this goal, a two-stage strategy combining LLM-guided program synthesis with evolutionary search is applied in the proposed method. In the first stage, a certain number of cost functions in the form of computer programs are synthesized from LLM responses to structured prompts. These cost functions are then evaluated with pretrained steganalysis models so that candidate cost functions suited to steganography can be collected. In the second stage, by retraining a steganalysis model for each candidate cost function, the optimal cost function(s) can be determined according to the detection accuracy. This two-stage strategy is performed by an iterative fashion so that the best cost function can be collected at the last iteration. Experiments show that the proposed method enables LLMs to design new cost functions of steganography that significantly outperform existing works in terms of resisting steganalysis tools, which verifies the superiority of the proposed method. To the best knowledge of the authors, this is the first work applying LLMs to the design of advanced cost function of steganography, which presents a novel perspective for steganography design and may shed light on further research.
\end{abstract}

\section{Introduction}
Steganography is referred to as the art of concealing secret information within a seemingly innocent cover to evade detection. Unlike cryptography, which scrambles the contents of a message to make it unreadable without a key, steganography hides the very existence of the message itself. This technique can be applied to various media formats, such as embedding secret text within digital images, audio signals, or video streams. The goal is to allow covert communication where only the intended recipient is aware of the hidden content. As digital communication becomes more prevalent, steganography plays an increasingly important role in fields such as cybersecurity, forensics, and privacy protection.

Among various types of cover media used in steganography, such as video, audio and text, images stand out as the most widely used and preferred option due to their high redundancy, broad format availability, ease of manipulation, and the human visual system’s low sensitivity to slight pixel-level changes, making them ideal for concealing information without raising suspicion. Moreover, many techniques originally designed for image steganography have the potential to be adapted and extended to other forms of cover media. A number of image steganography methods have been reported in the literature. Early methods are largely founded on empirical designs, aiming primarily to minimize the degree of alterations introduced by steganography. To reduce visual artifacts and enhance resistance against detectors based on handcrafted statistical features, adaptive steganography methods are introduced. These methods reveal the fact that different cover pixels have different suitability for steganography, which inspires the research community to propose the minimum-distortion embedding framework for steganography \cite{Pevny:IH:2010, Holub:EJIS:2014, Holub:WIFS:2012, Li:ICIP:2014}. In the framework, a cost function is defined over pixels in terms of statistical detectability or texture characteristics, where embedding data into a complex pixel generally gives a lower cost than a smooth one. Then, by applying syndrome-trellis codes (STCs) \cite{Filler:TIFS:2010} or other coding techniques, a stego image with the minimum embedding impact can be generated, which demonstrates superior performance in resisting various steganalyzers \cite{Fridrich:TIFS:2012, Xu:SPL:2016, Xu:IH:2016, Boroumand:TIFS:2018}. Many advanced methods are based on minimum-distortion embedding.

There are two widely applied strategies for cost function definition in above minimum-distortion embedding framework. One is manually crafted, and the other is based on deep learning. The former relies heavily on the designer's professional knowledge, whereas the latter learns the cost function through training a deep neural network, and therefore requires large-scale datasets. Generally, manually crafted methods \cite{Holub:EJIS:2014, Holub:WIFS:2012, Li:ICIP:2014} tend to utilize statistical features such as residual and context characteristics to construct a specific distortion function of image steganography. The advancement comes from the authors' expertise in steganography. Deep learning based techniques, in contrast, rely relatively less on human experience, but require a large amount of data for training a neural network that can generate a cost map or a probability map used to define the steganographic cost. Many advanced methods have been proposed in this direction. In summary, the evolution of cost function design has progressed through two distinct generations: experience-based and learning-based methods. 

In which direction will the next generation of cost function design head? We argue that the next generation may be reasoning-based. On the one hand, compared with experience-based methods, reasoning-based techniques primarily rely on machine intelligence, rather than relying entirely on the human expertise. On the other hand, compared with learning-based methods, reasoning-based techniques place greater emphasis on the model's ability to perform human-like reasoning, rather than the process of learning steganography-specific knowledge from large-scale data. Therefore, reasoning-based steganography significantly reduces the reliance on both expert knowledge and large-scale data. This shift marks a critical step towards achieving fully artificial intelligence (AI)-driven steganography design. A natural question that arises is therefore how reasoning-based steganography can be achieved. 

We believe that large language models (LLMs) may be a possible channel for solutions. This has already been confirmed by a prior study, although the research was relatively preliminary \cite{Wu:EI:2024}. LLMs have been receiving increasing attention in recent years, gaining widespread recognition for their remarkable capabilities \cite{Minaee:arXiv:2024}. LLMs have achieved great success across various domains such as natural language processing, machine translation, content generation, and human-computer interaction. This growing interest reflects both their practical impact and their potential to transform how we interact with information technology. LLMs exhibit strong ability to understand natural language and solve complex problems through text generation. This reasoning ability can be enhanced by prompt optimization, whose goal is to optimize the prompts fed into an LLM so that the LLM can return better solutions. Steganographic tasks can essentially be regarded as mathematical optimization problems. Although such problems may be deterministic or heuristic in nature, we posit that LLMs are well positioned to play a significant role in addressing them.

Unlike the work in \cite{Wu:EI:2024} where both the carrier and the secret information are presented explicitly within the prompt, we employ LLMs to generate computer programs capable of facilitating steganographic tasks. We aim to enable large models to discover novel steganographic cost functions in the form of computer programs. In detail, a two-stage strategy combining LLM-guided program synthesis with evolutionary search is proposed in this paper. The first stage generates a certain number of candidate cost functions. The second stage determines the present optimal cost function(s). These two stages are performed by an iterative fashion so that the best cost function is determined at the last iteration. Experimental results show that the proposed method enables LLMs to design new cost functions of steganography that significantly outperform existing works in terms of resisting steganalysis tools, which verifies the superiority and feasibility. This work provides a novel insight into the algorithmic design of steganography and has the potential to guide further research.

The rest structure of this paper is organized as follows. First of all, we provide preliminary concepts in next section. Then, we describe the proposed method in detail, followed by experimental results. Finally, we conclude this paper and provide discussions.

\begin{figure*}[!t]
	\centering
	\includegraphics[width=\linewidth]{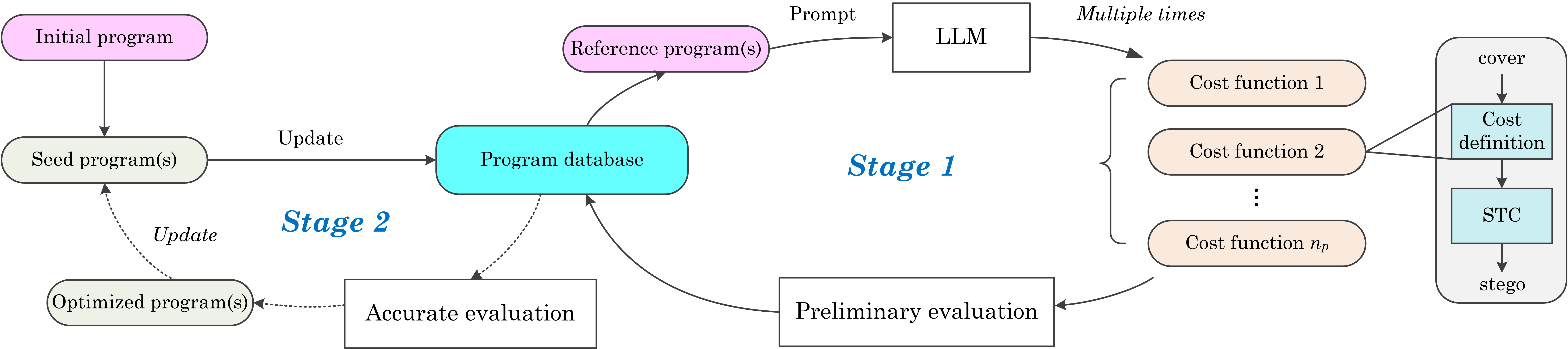}
	\caption{General framework for the proposed method, which contains two important stages. Stage 1 generates a certain number of candidate cost functions suited to steganography. Stage 2 determines the current optimal cost function(s). These two stages are performed by an iterative fashion.}\label{framework}
\end{figure*}

\section{Preliminary Concepts}
In this section, we provide necessary preliminary concepts in order to better introduce the proposed method later. 

\subsection{Large Language Models (LLMs)}
LLMs are advanced artificial intelligence systems trained on vast corpora of textual data to learn statistical patterns of natural language \cite{Minaee:arXiv:2024}. By using deep neural architectures, LLMs are capable of generating coherent, contextually relevant text, as well as performing a wide range of language-related tasks such as translation, summarization, and question answering. Their scalability in terms of both data and parameters has led to significant improvements in natural language understanding and generation, making them central to research in natural language processing (NLP).

The architecture underlying most LLMs is derived from the Transformer \cite{Vaswani:Transformer:2017}, which abandons recurrence and convolution in favor of attention mechanisms to model dependencies across sequence positions. Transformer‐based models are typically organized in one of three paradigms including encoder-only, decoder-only, or encoder-decoder. Encoder‐only architectures, e.g., BERT \cite{Devlin:BERT:2018}, are optimized for bidirectional context modeling and excel at natural language understanding tasks. Decoder‐only architectures such as GPT‐3 \cite{Brown:GPT3:2020}, are trained autoregressively to predict the next token and are suited for generative applications. Encoder-decoder architectures such as T5 \cite{Raffel:T5:2019}, leverage both encoder representations and decoder autoregression, enabling strong performance in text‐to‐text tasks such as translation, summarization, and question answering. In LLMs, the decoder-only architecture is the most common, but in its full form, the Transformer has both an encoder stack and a decoder stack.

LLMs have achieved remarkable success across many fields, with their application in solving complex mathematical problems being particularly noteworthy. These models can interpret natural language descriptions of real-world scenarios and automatically formulate them as optimization models, such as linear, nonlinear, or integer programs. They also assist in designing solution strategies, suggesting heuristics, and enhancing previous algorithms. It is especially useful when exact solutions are computationally expensive \cite{Jiang:ICLR:2025}. Moreover, LLMs can provide intuitive explanations of optimization results, making the outcomes more accessible to decision-makers in fields such as energy systems and finance \cite{Yang:ICLR:2024}. By bridging human reasoning and computational optimization techniques, LLMs are transforming both the modeling and solution processes in operations research and applied mathematics \cite{LLM:Nature:2024, Sartori:arXiv:2025}, inspiring us to explore LLMs for steganography.

\subsection{Steganography}
In minimum-distortion embedding framework \cite{Filler:TIFS:2010}, the cover sequence is denoted by $\textbf{x} = \{x_1, x_2, ..., x_n\}$, where $n$ represents the total number of cover elements and each cover element $x_i$ takes a value from a set $I$. By making slight modifications to these cover elements, we obtain a stego sequence $\textbf{y} = \{y_1, y_2, ..., y_n\}$. Each $y_i$ will be constrained within the neighborhood of $x_i$, i.e., $y_i \in N_i(x_i)$, e.g., $N_i(x_i) = \{x_i - 1, x_i, x_i + 1\}$ for ternary embedding (also called $\pm 1$ embedding). Without the loss of generalization, we will consider $\textbf{x}$ as a 8-bit grayscale image, i.e., $I = \{0, 1, ..., 255\}$.

The impact of modifications caused by steganography can be measured by a distortion function $D(\textbf{x}, \textbf{y})$, which is expressed as:
\begin{equation}
D(\mathbf{x},\mathbf{y}) = \sum_{i=1}^n \rho_i(\textbf{x},y_i),
\end{equation}
where $\rho_i(\textbf{x},y_i)$ quantifies the cost incurred by modifying $x_i$ to $y_i$. Here, we restrict the distortion function to an additive form, i.e., the individual costs $\rho_i(\textbf{x},y_i)$, $1\leq i\leq n$, are independent of each other. Given a constraint on the payload, the sender aims to minimize the expected total distortion subject to a fixed embedding rate of $m$ bits. This optimization problem can be formalized as:
\begin{equation}
E_\pi[D] = \sum_{\textbf{y} \in \prod_{i=1}^{n}N_i(x_i)} \pi(\textbf{y})D(\textbf{x}, \textbf{y})
\end{equation}
subject to
\begin{equation}
H(\pi) = m = \alpha n,
\end{equation}
where $\prod_{i=1}^{n}N_i(x_i)$ represents the set of all stego sequences, $\pi$ is their probability distribution characterizing the steganographer's actions, and $\alpha$ gives the hidden bits per pixel (bpp). The optimal distribution $\pi$ for the problem has the Gibbs form \cite{Filler:Gibbs:2010}, i.e.,
\begin{equation}
\pi_\lambda(\textbf{y}) = \frac{e^{-\lambda D(\textbf{x}, \textbf{y})}}{\sum_{\textbf{y}' \in \prod_{i=1}^{n}N_i(x_i)}e^{-\lambda D(\textbf{x}, \textbf{y}')}}
\end{equation}
where $\lambda$ is a parameter used to satisfy the payload constraint.

Under the assumption that changing a pixel always results in an increased distortion, for binary embedding for which we have $N_i(x_i) = \{x_i, \bar{x}_i\}$, $1\leq i\leq n$, one can rewrite Eq. (1) as
\begin{equation}
D(\textbf{x}, \textbf{y}) = \sum_{i=1}^{n}\epsilon_i\cdot[y_i \neq x_i],
\end{equation}
where $\epsilon_i = |\rho_i(\textbf{x}, x_i) - \rho_i(\textbf{x}, \bar{x}_i)| > 0$ and $[\cdot]$ is the indicator function. Accordingly, the optimal probability distribution requires us to modify each cover pixel with
probability
\begin{equation}
\pi_i = \frac{e^{-\lambda \epsilon_i}}{1 + e^{-\lambda \epsilon_i}}, 1\leq i\leq n.
\end{equation} 

For ternary embedding, one may assume $\rho_i(\textbf{x}, x_i) = 0$ and $\rho_i(\textbf{x}, x_i - 1) = \rho_i(\textbf{x}, x_i + 1) = \epsilon_i' > 0$. The overall distortion can be expressed as the form similar to Eq. (5), but the optimal distribution requires us to modify each cover pixel with probability
\begin{equation}
\pi_i = \frac{2e^{-\lambda \epsilon_i'}}{1 + 2e^{-\lambda \epsilon_i'}}, 1\leq i\leq n.
\end{equation}

It is free to apply other embedding operations. However, regardless of the embedding operation that we use, under the above minimum-distortion embedding framework, the most important task is to evaluate the cost of modifying each pixel, i.e., $\epsilon_i$ and $\epsilon_i'$. Once the cost for each cover pixel can be well measured, practical codes such as STCs \cite{Filler:TIFS:2010} can be applied to approach the performance. This paper aims to leverage LLMs to quantify the cost.

\subsection{Steganalysis}
The evolution of (image) steganography and steganalysis has long been mutually driven, with each advancing in response to the other. Steganography aims to enhance the imperceptibility of embedded information, while steganalysis strives to uncover such concealed manipulations. Steganalysis can be modeled as a binary classification problem that seeks to differentiate between visually indistinguishable cover and stego images, thereby detecting potential covert communication. Because embedding operations inevitably alter an image’s statistical characteristics, even minute modifications can leave discernible traces in pixel correlations or high-frequency components. Steganalysis identifies these subtle artifacts to expose the presence of hidden information.

Traditional steganalysis methods rely on handcrafted feature extractors and classifiers. These approaches typically construct image residuals or transition probability matrices to capture statistical features, which are then classified using machine learning algorithms such as support vector machine (SVM). Early steganalysis techniques were often designed for specific steganographic schemes, leveraging prior knowledge to craft discriminative, task-oriented features. Subsequent advancements introduced multi-scale and multi-directional residual models, extracting high-dimensional representations from hundreds of high-pass filters to improve generalization. Nevertheless, their dependence on manually designed features and the separate optimization of feature extraction and classification stages restrict their adaptability under complex real-world conditions.

Recently, research on steganalysis has undergone rapid advancements driven by deep learning techniques \cite{Xu:SPL:2016, Xu:IH:2016, Boroumand:TIFS:2018}. Basically, given the input image, it is firstly preprocessed using high-pass filters or residual layers to amplify steganographic signals, allowing deep convolutional neural networks (CNNs) to learn the subtle artifacts introduced by embedding operations. The final fully connected layers then perform classification between cover and stego images. Building upon this foundation, subsequent models have refined network architectures and training strategies, leading to continuous improvements in detection accuracy, robustness, and generalization. A notable milestone is SRNet \cite{Boroumand:TIFS:2018}, which unifies feature extraction and classification within an end-to-end learning framework. Unlike earlier CNN-based methods that relied on fixed handcrafted filters, SRNet allows the network to learn effective filter kernels directly from data, facilitating the autonomous extraction of discriminative steganographic features. Owing to its stable feature representation and strong generalization across various steganographic algorithms, SRNet has become a widely adopted baseline model. Consequently, this study uses SRNet as the performance evaluator for steganographic schemes.

\section{Proposed Method}
To automate the design of steganographic algorithms, we introduce a novel two-stage framework that integrates LLMs with an evolutionary (code) search mechanism. The proposed framework evolves over multiple generations, each comprising two important stages, namely, cost function generation and cost function evaluation. As shown in Figure \ref{framework}, Stage 1 employs prompt-guided LLMs to iteratively generate and refine code for cost functions, which are subjected to rapid preliminary evaluation. Stage 2 conducts a more rigorous and accurate assessment of the cost functions produced in Stage 1 and provides targeted feedback for refinement. Through repeated evolution across generations, this framework accumulates a diverse set of candidate functions and progressively identifies superior cost functions under robust evaluation. Different from conventional, manually designed heuristic methods, the proposed framework enables automated exploration of expansive search spaces and continuously adjusts its search trajectory through feedback, thereby substantially enhancing the efficiency and creativity of algorithmic design. We show more details about the two stages in the following subsections.

\subsection{Stage 1: Cost Function Generation}
The objective of Stage 1 is to efficiently identify and preliminarily filter a set of candidate cost functions, ensuring both the scalability and search efficiency of the framework. To achieve this objective, an iterative evolution-based search strategy driven by LLM is employed. With structured prompts, the LLM iteratively refines existing cost functions to produce multiple improved variants. These variants then undergo rapid evaluation, and those that execute successfully and yield valid scores are incorporated into the program database as candidate cost functions for steganography. All cost functions are in the form of computer programs. We describe the workflow of Stage 1 in the following.

Mathematically, a classical heuristic cost function, expressed as $p_\text{raw}$, is selected as the initial program $p_\text{init}$ at the very beginning, which is then stored as the seed program $p_\text{seed}$ and incorporated into the program database. Such a database can be expressed as $S_\text{P} = \{\text{PD}_1, \text{PD}_2, ..., \text{PD}_n\}$, $n > 0$, where $\text{PD}_i$ is a sub-database. Here, $S_\text{P}$ and $\text{PD}_i~(1\leq i\leq n)$ are all sets. It is noted that the value of $n$ should be determined in advance. Initially, we have
\begin{equation}
\text{PD}_1 = \text{PD}_2 = ... = \text{PD}_n = \emptyset.
\end{equation}
By adding $p_\text{seed} = p_\text{init} = p_\text{raw}$ to $S_\text{P}$, we have 
\begin{equation}
	\text{PD}_1 = \text{PD}_2 = ... = \text{PD}_n = \{\{p_\text{raw}\}\}.
\end{equation}
In other words, the initial program is present in each sub-database. In fact, each $\text{PD}_i$ consists of multiple clusters of programs, which will be clarified later. Obviously, each $\text{PD}_i$ now consists of only one cluster, i.e., $\{p_\text{raw}\}$. A reference program, denoted by $p_\text{ref}$, is thereafter sampled from $S_\text{P}$ as follows: First, an index $1\leq i\leq n$ is randomly generated; Then, $p_\text{ref}$ is sampled from some cluster of $\text{PD}_i$ based on the temperature-scaled softmax function used to determine the selection probability of each cluster. Clearly, suppose that $\text{PD}_i = \{\text{C}_{i,1}, \text{C}_{i,2}, ..., \text{C}_{i,k_i}\}$, where $k_i > 0$ is the total number of clusters in $\text{PD}_i$. Each $\text{C}_{i,j}$ is associated with a score $s_{i,j}$. Here, a larger $s_{i,j}$ implies that the cost functions in $\text{C}_{i,j}$ have better performance resisting against the steganalysis tools. We will provide more information about $s_{i,j}$ later. In this way, the probability of sampling $\text{C}_{i,j}$ from $\text{PD}_i$ is given by
\begin{equation}
\text{Pr}\{\text{C}_{i,j}~|~\text{PD}_i\} = \frac{e^{s_{i,j}/T}}{\sum_{r=1}^{k_i}e^{s_{i,r}/T}},
\end{equation}
where the temperature parameter $T$ regulates the balance between exploration and exploitation. A larger $T$ introduces more randomness by allowing low-scoring clusters to be occasionally chosen, while a smaller $T$ focuses the search on high-scoring clusters. For the case that there is only one program in $\text{C}_{i,j}$, the program will be selected as $p_\text{ref}$. Otherwise, one may randomly select one program as $p_\text{ref}$. In our simulation, we use another softmax function to sample the program. Let $\text{C}_{i,j} = \{c_{1}, c_{2}, ..., c_{|\text{C}_{i,j}|}\}$, where $c_k$ is the $k$-th program. The length of $c_k$, i.e., the number of characters, is denoted by $l_k$. For all $1\leq k\leq |\text{C}_{i,j}|$, we normalize $l_k$ by
\begin{equation}
l_k' = \frac{l_k - \min\{l_1, l_2, ..., l_{|\text{C}_{i,j}|}\}}{\max\{l_1, l_2, ..., l_{|\text{C}_{i,j}|}\} - \min\{l_1, l_2, ..., l_{|\text{C}_{i,j}|}\}+\epsilon},
\end{equation}
where $\epsilon = 10^{-6}$. Obviously, $l_k'\in [0,1),~\forall 1\leq k\leq |\text{C}_{i,j}|$. Accordingly, the probability of selecting $c_k$ as $p_\text{ref}$ in our simulation is
\begin{equation}
\text{Pr}\{c_k~|~\text{C}_{i,j}\} = \frac{e^{-l_k'}}{\sum_{r=1}^{|\text{C}_{i,j}|}e^{-l_r'}},~\forall 1\leq k\leq |\text{C}_{i,j}|,
\end{equation}
which suggests a shorter program as $p_\text{ref}$ as longer programs are more prone to errors in terms of both computer programming and cost function definition.

After sampling the reference program $p_\text{ref}$ out from the program database, we are ready to construct a prompt to be fed into the LLM for cost function discovery. This prompt consists of two components, i.e., $p_\text{ref}$ and a natural language instruction. The former provides an executable example for mutation, while the latter specifies the intended evolutionary direction. A placeholder function is reserved for evolution, sharing the same input and output parameters as $p_\text{ref}$, but with a different function name. The body of the placeholder function will be generated by the LLM. The natural language prompt defines the generation objectives, design constraints, and output requirements, while also encouraging the generation of more complex and innovative cost function. In fact, we can further extend the above operation to a more general case. 

Specifically, we sample multiple reference programs (if any) from the program database, denoted by $p_\text{ref}^{(0)}$, $p_\text{ref}^{(1)}$, ..., $p_\text{ref}^{(r-1)}$. The above operation corresponds to $r\equiv 1$. We assign a version identifier by appending a suffix to the cost function name of each reference program. For example, if the original name of cost function for $p_\text{ref}^{(0)}$ is $\texttt{compute\_cost\_adjusted}$, it will be then modified as $\texttt{compute\_cost\_adjusted\_v0}$. Thus, the LLM aims to generate the body for the function $\texttt{compute\_cost\_adjusted\_vr}$, where $\texttt{r}$ will be replaced by a number. In experiments, we set $r\equiv 1$. It is noted that $p_\text{ref}^{(0)}$, $p_\text{ref}^{(1)}$, ..., $p_\text{ref}^{(r-1)}$ come from the same sub-database.

By feeding the same prompt into the LLM $n_p$ times, we can obtain $n_p$ evolved candidate functions. All $n_p$ evolved functions are subjected to the preliminary evaluation, which filters out non-executable candidates and assigns coarse scores to the executable ones. The evaluation procedure employs a set of pretrained deep steganalysis models for assessment. We adopt SRNet \cite{Boroumand:TIFS:2018} as the architecture in this paper. Initially, the evaluation pool consists of multiple SRNet models trained using different heuristic steganographic algorithms. Each SRNet model enables us to evaluate a program by the minimum average decision error rate \cite{Filler:TIFS:2010}, i.e.,
\begin{equation}
	P_\text{E} = \text{min}_{P_\text{FA}}~\frac{1}{2}\left(P_\text{FA}+P_\text{MD}(P_\text{FA})\right),
\end{equation}
where $P_\text{FA}$ and $P_\text{MD}$ denote the false-alarm and missed-detection rates, respectively. The evaluation score of an evolved program is defined as the average $P_\text{E}$ over all the steganalysis models in the evaluation pool. Only executable programs (functions) are associated with an evaluation score. Non-executable ones are discarded. Moreover, this evaluation score corresponds to the variable mentioned in Eq. (10). Without the loss of generalization, let $E_1$, $E_2$, ..., $E_{n_q}$ denote all the executable programs each corresponding to a cost function. $s(E_k)$ represents the evaluation score of $E_k$. Obviously, $n_q \leq n_p$. Each $E_k$ will be inserted into the sub-database corresponding to $p_\text{ref}$, implying that $E_k$ was derived from $p_\text{ref}$. And, $E_k$ will be added to the cluster of the sub-database whose score is equal to $s(E_k)$. If such cluster does not exist, a new cluster containing $E_k$ will be created, and its score is $s(E_k)$. It is noted that the score of $p_\text{raw}$ in Eq. (9) is determined by the same way.

The above operations will be performed in an iterative manner, resulting in an increasing number of programs in the database. To improve efficiency, after performing the above operations multiple times, we remove `bad' programs from the database. First, all the sub-databases are marked as `good' or `bad' based on their highest cluster scores. That is, for each sub-database, if its highest cluster score is no less than a threshold, it is marked as `good', otherwise it is marked as `bad'. Then, each `bad' sub-database is replaced with a new sub-database that only contains the highest-score cluster of a randomly selected `good' sub-database. Thus, all sub-databases become `good', i.e., most programs are `good'. For implementation, it may be not easy to specify the value of the threshold. An alternative strategy is to set the percentage of `bad' sub-databases to be replaced, e.g., 50\%, which has been applied in our experiments. In conclusion, it is always free for us to design the strategy updating the database so as to improve efficiency.

\emph{Remark:} As mentioned previously, we use pretrained deep steganalysis models for preliminary evaluation. These steganalysis models were pretrained with a fixed embedding payload, e.g., 0.4 bits per pixel (bpp) in our experiments. This guarantees that the assessment of evolved programs reflects their security performance under a standardized condition, preventing the evolutionary search from exploiting the detection characteristics of a varying embedding rate. The generalization of the evolved functions will, in a later phase, be validated via a multi-rate assessment.

\subsection{Stage 2: Cost Function Evaluation}
Once the code generation process reaches the preset iteration limit, the system proceeds to Stage 2, i.e., the accurate evaluation phase. At this point, all programs except the seed program are extracted from every cluster across all sub-databases in the program database to form the candidate program set $\text{PS}$. If the total number of programs in $\text{PS}$ is less than a threshold, all the programs in $\text{PS}$ will undergo accurate evaluation, otherwise 50\% programs are randomly selected for accurate evaluation to improve efficiency.

Each program corresponds to a cost function, which can be used to train a dedicated steganalyzer, using stegos generated by the cost function and the corresponding covers. The accurate evaluation phase in this paper uses SRNet as the network architecture. Accordingly, each program corresponds to an accurate evaluation score, i.e., $P_\text{E}$ shown in Eq. (13). We collect all programs whose accurate evaluation scores are no less than a threshold $s_\text{init} - \epsilon$, in which $s_\text{init}$ represents the accurate evaluation score of $p_\text{init}$ and $\epsilon$ is a small threshold. These programs are called optimized programs as shown in Figure \ref{framework}, which are  merged into the seed program set, serving as the initialization basis for the subsequent evolutionary cycle. The program database is then updated by clearing its contents and repopulating each sub-database with all seed programs, where their scores are reassessed using the updated preliminary evaluation set to form a new generation of candidates. 

Additionally, for any program (cost function) that exhibits a high preliminary evaluation score but performs poorly in accurate evaluation, it indicates that its steganographic features are inadequately covered by the preliminary evaluation set. In this case, the steganalyzer corresponding to that program will be incorporated into the preliminary evaluator collection. Therefore, through this mechanism, the system progressively strengthens the discriminative capability of the preliminary evaluation stage for diverse candidate programs, effectively enhancing the reliability and stability of the entire evolutionary evaluation process.

\emph{Remark:} After multiple rounds of evolution, programs that achieve good performance in both preliminary and accurate evaluation, are marked as potential functions. Obviously, these potential functions are all in the program database. They are then evaluated across different embedding rates. For each of these candidate cost functions, we train steganalyzers using different embedding rates and perform accurate evaluation for each embedding rate. A candidate is only accepted as a superior cost function if it outperforms the initial program across all embedding rates. This multi-rate validation is essential because it enforces cross-rate generalization and prevents overfitting to a single rate, ensuring that the final promoted functions reflect genuine and robust security improvements. If there are multiple superior cost functions, any one can be selected as the final cost function generated by the LLM. One may also design a new strategy to select one of them.

\section{Experimental Results and Analysis}
In this section, we provide experiments and analysis to verify the feasibility and superiority of the proposed work.

\subsection{Setup}
We conduct extensive experiments to validate the effectiveness of the proposed method. The LLM used for code generation in our experiments is OpenAI GPT-4o \cite{GPT4o:arXiv:2024}, which, as one of the most widely adopted and recent LLM releases, provides strong performance in instruction following and structured output generation. It delivers fast and reliable responses while maintaining a relatively cost-effective API pricing structure. The steganalysis models in Stage 1 are pretrained at an embedding rate of 0.4 bpp, and the heuristic steganographic algorithms are WOW \cite{Holub:WIFS:2012}, HILL \cite{Li:ICIP:2014} and SUNIWARD \cite{Holub:EJIS:2014}. In other words, the evaluation pool in Stage 1 contains three SRNet \cite{Boroumand:TIFS:2018} models initially. These three cost functions are used as the initial program for independent evolution. For steganalysis, the BOSSBase v1.01 dataset \cite{Bas:IH:2011} is used, in which all the images are resized into $256\times 256$ due to limited computing resources. The ratio between training set, validation set and test set is 5:1:4. Python is used for program development. The embedding simulator is used for implementing all algorithms.

The number of sub-databases is set to two, i.e., $|\text{S}_\text{P}| = n = 2$. The temperature parameter $T$ in Eq. (10) is set to 0.1 initially, and then gradually decreases as the number of iterations increases. In each iteration, the framework samples only one reference function (i.e., $r = 1$), which is then combined with a fixed natural language prompt. The resulting prompt is therefore fed into the LLM four times to generate four new evolved candidate functions, i.e., $n_p = 4$. After an initial burn-in period of ten iterations, the process enters Stage 2 for accurate evaluation, subsequently re-entering Stage 2 every five iterations, ensuring frequent assessment to effectively guide the search. To avoid long-term stagnation caused by low-quality programs, the program database is refreshed every four hours (here, we use the running time to determine whether to update the program database) by removing `bad' programs based on the preliminary evaluation scores of corresponding functions.

In Stage 2, the threshold for the number of programs selected for accurate evaluation is set as 10, meaning that if the total number of programs in $\text{PS}$ is less than 10, all the programs will undergo accurate evaluation, otherwise 50\% programs are randomly selected for accurate evaluation. An optimized program is designated only if its accurate evaluation score is no less than the score of the initial program, i.e., $s_\text{init}$, by a margin $\epsilon = 1\%$. For the potential candidate function, to ensure the reliability of the performance improvement, we convert the evolved Python programs into MATLAB implementations. These MATLAB programs are then subjected to a rigorous multi-rate assessment at 0.4 bpp, 0.3 bpp, 0.2 bpp, and 0.1 bpp, and compared against their corresponding initial programs. We hope to generate three new cost functions outperforming WOW, HILL and SUNIWARD, respectively.

For SRNet training at 0.4 bpp, the initial learning rate is set as $10^{-3}$, and reduced to $10^{-4}$ if the validation accuracy does not improve for 100 consecutive epochs. Training is terminated early if no improvement is observed after 200 epochs, with a maximum of 800 epochs. All remaining training settings are consistent with those described in the original paper \cite{Boroumand:TIFS:2018}. For training SRNet at other embedding rates, we use a progressive fine-tuning strategy: the detector at 0.3 bpp is initialized from the trained model at 0.4 bpp, the detector at 0.2 bpp is initialized from that at 0.3 bpp, and so on. Each stage involves 100 epochs of training at a learning rate of $10^{-3}$, followed by 50 epochs at $10^{-4}$. The best validation snapshot from the final 50 epochs is selected as the final detector.

\subsection{Evolved Cost Functions}
We use the cost functions of WOW, HILL and SUNIWARD as initial functions, and then respectively produce three superior cost functions with enhanced ability resisting steganalysis. Below, we describe each of the evolved cost functions in detail.

\emph{1) Cost function evolved from WOW:} Before we report the evolved algorithm, we first describe the original WOW algorithm \cite{Holub:WIFS:2012}. WOW defines its cost function using a Daubechies 8-tap (DB-8) wavelet directional filter bank. This bank, denoted by $\textbf{K}^{(k)}$, $k = 0, 1, 2$, is composed of the LH, HL, and HH directional high-pass filters, expressed as $\textbf{K}^{(0)} = \textbf{h}\cdot \textbf{g}^\text{T}$, $\textbf{K}^{(1)} = \textbf{g}\cdot \textbf{h}^\text{T}$, and $\textbf{K}^{(2)} = \textbf{g}\cdot \textbf{g}^\text{T}$, where $\textbf{h}$ and $\textbf{g}$ denote the Daubechies-8 decomposition low-pass and high-pass filters, respectively. 

The embedding suitability, i.e., $\xi^{(k)}$, quantifies the local texture complexity through measuring the weighted absolute differences between the filter residuals of a cover image and those after modifying the value of a single pixel to the target value. Namely,
\begin{equation}
\xi^{(k)} = \left| \textbf{X} \otimes \textbf{K}^{(k)} \right| \odot \left|\textbf{K}^{(k)}\right|,~k = 0, 1, 2,
\end{equation} 
where $\otimes$ denotes the mirror-padded convolution and $\odot$ represents the mirror-padded correlation. Here, $\textbf{X}$ represents the cover image sized $h\times w$, where $h$ and $w$ denote the height and width.

The formulation presented here is derived from the observation that the residual differences correspond to rotated directional filters, with the absolute values of filter residuals serving as adaptive weights. The underlying principle is that the pixels exhibiting small filter residuals along any directional component are highly predictable within their local context. Consequently, such pixels are assigned higher embedding costs to minimize the detectability. The final embedding cost $\rho$ is determined by aggregating the reciprocal of all suitability measures, i.e.,
\begin{equation}
\rho = \frac{1}{\xi^{(0)}} + \frac{1}{\xi^{(1)}} + \frac{1}{\xi^{(2)}}.
\end{equation}

This ensures that predictable regions (with small residuals) receive higher costs, while complex textures (with large residuals) are favored for embedding. The WOW-evolved algorithm extends WOW through four principal innovations, including 
\begin{itemize}
	\item[P1:] \emph{An expanded directional filter bank.}
	\item[P2:] \emph{Gaussian smoothing of individual filters.}
	\item[P3:] \emph{Sensitivity-aware weighting mechanisms.}
	\item[P4:] \emph{A comprehensive suite of post-processing operations.}
\end{itemize}

Specifically, it captures high-frequency details across various directions by using a set of five convolutional kernels, expressed as $\textbf{K}_\text{e}^{(k)},~0\leq k\leq 4$, which are defined by two 1-D difference operators $\textbf{d}_0 = \frac{1}{2}(1, 1)^\text{T}$ and $\textbf{d}_1 = \frac{1}{2}(1, -1)^\text{T}$. These kernels detect vertical, horizontal, diagonal, and non-directional features, supplemented by a scaled identity matrix to amplify local variance:
\begin{equation}
\textbf{K}_\text{e}^{(0)} = \textbf{d}_1 \cdot \textbf{d}_0^\text{T} = \frac{1}{4}\begin{bmatrix}
		1 & 1\\
		-1 & -1
\end{bmatrix}
\end{equation}
\begin{equation}
	\textbf{K}_\text{e}^{(1)} = \textbf{d}_0 \cdot \textbf{d}_1^\text{T} = \frac{1}{4}\begin{bmatrix}
		1 & -1\\
		1 & -1
	\end{bmatrix}
\end{equation}
\begin{equation}
	\textbf{K}_\text{e}^{(2)} = \textbf{d}_1 \cdot \textbf{d}_1^\text{T} = \frac{1}{4}\begin{bmatrix}
		1 & -1\\
		-1 & 1
	\end{bmatrix}
\end{equation}
\begin{equation}
	\textbf{K}_\text{e}^{(3)} = \textbf{d}_0 \cdot \textbf{d}_0^\text{T} = \frac{1}{4}\begin{bmatrix}
		1 & 1\\
		1 & 1
	\end{bmatrix}
\end{equation}
\begin{equation}
	\textbf{K}_\text{e}^{(4)} = 2\textbf{I} = 
	\begin{bmatrix}
		2 & 0\\
		0 & 2
	\end{bmatrix}
\end{equation}
For each kernel $\textbf{K}_\text{e}^{(k)}$, an embedding suitability map $\xi_\text{e}^{(k)}$ is generated by
\begin{equation}
\xi_\text{e}^{(k)} = \left| \textbf{X} \otimes \textbf{K}_\text{e}^{(k)} \right| \otimes \textbf{G}_\sigma,~0\leq k\leq 4,
\end{equation}
where $\textbf{G}_\sigma$ is a Gaussian filter with $\sigma = 0.8$ and each element at $(i,j)$ with the center of $\textbf{G}_\sigma$ as the origin is $\textbf{G}_\sigma[i,j] = \frac{1}{2\pi\sigma^2}e^{-\frac{i^2+j^2}{2\sigma^2}}$. The discrete kernel size is defined as $(2\times \left \lceil  L\sigma - 0.5 \right \rceil + 1)^2$, which yields a $7\times 7$ kernel by setting $L = 4$ by default. 

The final embedding suitability map $\xi_\text{e}$ is determined by applying a custom sensitivity weight vector $\textbf{w}$ and a sensitivity factor $p$ to a power sum of the individual embedding suitability maps:
\begin{equation}
\xi_\text{e} = \sum_{k=0}^{4}w_k\left(\xi_\text{e}^{(k)}\right)^p = \sum_{k=0}^{4}w_k\left(\xi_\text{e}^{(k)}[i,j]\right)_{h\times w}^p,
\end{equation}
where $\textbf{w} = (1.8, 1.4, 1.6, 1.0, 0.9)$, $p = -2.5$, and $\xi_\text{e}^{(k)}[i,j]$ represents the element at position $(i,j)$ of $\xi_\text{e}^{(k)}$. We initialize a cost map as follows:
\begin{equation}
\rho_{i,j}' = \left\{\begin{matrix}
+\infty	& \text{if}~\xi_\text{e}[i,j]=0,\\
\left(\xi_\text{e}[i,j]\right)^{-1/p} &  \text{if}~\xi_\text{e}[i,j]\neq 0.
\end{matrix}\right.
\end{equation}
This cost is then refined as:
\begin{equation}
\rho_{i,j} = \left\{\begin{matrix}
	+\infty	& \text{if}~\rho_{i,j}' < \theta,\\
	\rho_{i,j}' &  \text{if}~\rho_{i,j}' \geq \theta,
\end{matrix}\right.
\end{equation}
where $\theta = 10^{-12}$. We use $\rho_{i,j}^{(+1)}$ and $\rho_{i,j}^{(-1)}$ to denote the cost by modifying the cover pixel $x_{i,j} \in \textbf{X}$ with `+1' and `-1' operations, respectively. The evolved algorithm initializes them as follows:
\begin{equation}
\rho_{i,j}^{(+1)} = \left\{\begin{matrix}
	+\infty	& \text{if}~x_{i,j} > 255 - \tau,\\
	\rho_{i,j} &  \text{if}~x_{i,j} \leq 255 - \tau,
\end{matrix}\right.
\end{equation}
and
\begin{equation}
\rho_{i,j}^{(-1)} = \left\{\begin{matrix}
	+\infty	& \text{if}~x_{i,j} < \tau,\\
	\rho_{i,j} &  \text{if}~x_{i,j} \geq \tau,
\end{matrix}\right.
\end{equation}
where $\tau = 1$ by default. $\tau$ prevents the algorithm from modifying boundary pixels. 

Finally, the evolved algorithm clamps the top 5\% of embedding costs to infinity to avoid modifications in the most vulnerable regions. Let $T_{5\%}^{(+1)}$ and $T_{5\%}^{(-1)}$ represent the thresholds defining the boundary of the top $5\%$ for `+1' and `-1' operations, respectively. The final cost maps are determined by
\begin{equation}
	\rho_{i,j}^{(+1)} = \left\{\begin{matrix}
		+\infty	& \text{if}~\rho_{i,j}^{(+1)} > T_{5\%}^{(+1)},\\
		\rho_{i,j}^{(+1)} &  \text{if}~\rho_{i,j}^{(+1)} \leq T_{5\%}^{(+1)},
	\end{matrix}\right.
\end{equation}
and
\begin{equation}
	\rho_{i,j}^{(-1)} = \left\{\begin{matrix}
		+\infty	& \text{if}~\rho_{i,j}^{(-1)} > T_{5\%}^{(-1)},\\
		\rho_{i,j}^{(-1)} &  \text{if}~\rho_{i,j}^{(+1)} \leq T_{5\%}^{(-1)}.
	\end{matrix}\right.
\end{equation}

\emph{2) Cost function evolved from HILL:} The original HILL algorithm adopts a $3\times 3$ Ker-Bohme (KB) filter as a high-pass filter $\textbf{H}_1$ to extract high-frequency image details, i.e.,
\begin{equation}
\textbf{H}_1 = \begin{bmatrix}
	-1  & 2 & -1\\
	2  & -4 & 2\\
	-1  & 2 & -1
\end{bmatrix}.
\end{equation}
With a $3\times 3$ average filter $\textbf{L}_1$, the embedding suitability map $\xi$ is generated by smoothing the absolute residuals:
\begin{equation}
\xi = \left| \textbf{X} \otimes \textbf{H}_1 \right| \otimes \textbf{L}_1.
\end{equation}
HILL then applies a $15\times 15$ average filter $\textbf{L}_2$ to smooth the reciprocal of the embedding suitability and compute the final embedding cost as
\begin{equation}
\rho = \frac{1}{\xi+\epsilon}\otimes \textbf{L}_2,
\end{equation}
where $\epsilon$ can be set to, e.g., $10^{-10}$.

The HILL-evolved algorithm improves HILL by simply replacing $\textbf{L}_2$ with a $25\times 25$ Gaussian filter $\textbf{G}_\sigma$ where $\sigma = 3$. Thus, the embedding cost of the HILL-evolved algorithm is given by
\begin{equation}
\rho_\text{e} = \frac{1}{\xi+\epsilon}\otimes \textbf{G}_\sigma.
\end{equation}

\emph{3) Cost function evolved from SUNIWARD:} Similar to WOW, the original SUNIWARD algorithm also adopts the DB-8 wavelet directional filter bank for cost definition. SUNIWARD first determines the residuals as follows:
\begin{equation}
\textbf{R}^{(k)} = \textbf{X}\otimes \textbf{K}^{(k)},~k = 0, 1, 2.
\end{equation}
Then, the embedding suitability, $\xi^{(k)}$, quantifies the local texture complexity by
\begin{equation}
\xi^{(k)} = \left( \frac{1}{|\textbf{R}^{(k)}| + \epsilon} \right)\odot \left|\textbf{K}^{(k)}\right|,~k = 0, 1, 2,
\end{equation}
where $\epsilon$ is set to 1. The final embedding cost is then given by
\begin{equation}
\rho = \xi^{(0)} + \xi^{(1)} + \xi^{(2)},
\end{equation}
which suppresses modifications in smooth, predictable regions by penalizing changes to small coefficients, while tolerating them in complex textures where large coefficients can absorb distortions. 

The SUNIWARD-evolved algorithm introduces two key enhancements to the original SUNIWARD algorithm, i.e.,
\begin{itemize}
	\item[P1:] \emph{Incorporation of directional weighting to emphasize specific filter orientations.}
	\item[P2:] \emph{Introduction of boundary-aware processing.}
\end{itemize}

Directional sensitivity is reinforced by two-stage weighting. In the first stage, the residual differences are processed with a directional weighting factor during the computation of the evolved suitability map, which can be expressed as:
\begin{equation}
\xi_\text{e}^{(k)} = \left( \frac{1}{|\textbf{R}^{(k)}| + \epsilon} \right)\odot \left(\alpha_k\left|\textbf{K}^{(k)}\right|\right),~k = 0, 1, 2,
\end{equation}
where $\alpha = (1, 1.5, 1)$. This operation amplifies the sensitivity to vertical structural details. Then, in the second stage, the cost map is obtained by using a weighted summation of the suitability maps instead of a direct sum, i.e.,
\begin{equation}
\rho_\text{e} = \beta_0\xi_\text{e}^{(0)} + \beta_1\xi_\text{e}^{(1)} + \beta_2\xi_\text{e}^{(2)},
\end{equation}
where $\beta = (0.5, 1, 0.5)$, further accentuating the contribution of the vertical directional component. To enhance security in boundary regions, the algorithm introduces a threshold $\tau = 5$, preventing information embedding in extremely bright or dark areas, i.e.,
\begin{equation}
	\rho_{i,j}^{(+1)} = \left\{\begin{matrix}
		+\infty	& \text{if}~x_{i,j} \geq 255 - \tau,\\
		\rho_{i,j} &  \text{if}~x_{i,j} < 255 - \tau,
	\end{matrix}\right.
\end{equation}
and
\begin{equation}
	\rho_{i,j}^{(-1)} = \left\{\begin{matrix}
		+\infty	& \text{if}~x_{i,j} \leq \tau,\\
		\rho_{i,j} &  \text{if}~x_{i,j} > \tau.
	\end{matrix}\right.
\end{equation}

\begin{figure*}[!t]
	\centering
	\includegraphics[width=\linewidth]{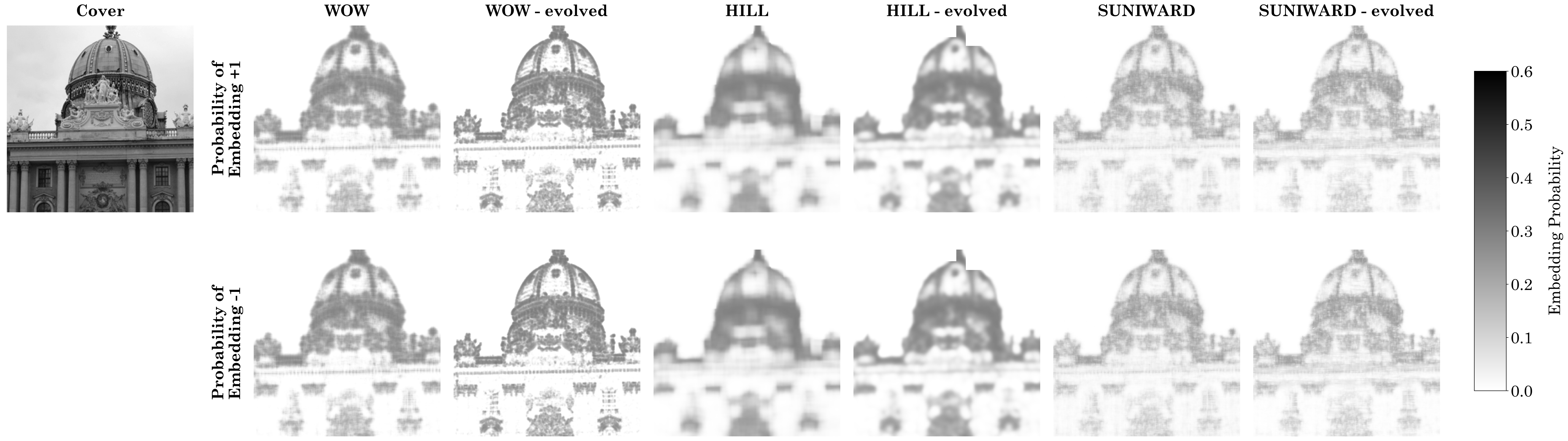}
	\caption{The embedding probability maps at an embedding rate of 0.4 bpp by applying different cost functions to a grayscale cover with a size of $256\times 256$.}\label{probMap}
\end{figure*}

\begin{table}[!t]
	\caption{Table 1. Performance comparison between the evolved algorithms and their original versions (round to 4 decimal places).}
	\begin{center}
		\begin{tabular}{c|cccc}
			\hline\hline
			\multirow{2}{*}{Algorithm} & \multicolumn{4}{c}{Embedding rate (bpp)} \\
			& 0.1 & 0.2 & 0.3 & 0.4\\
			\hline
			WOW & .2640 & .1800 & .1361 & .1053\\
			WOW-evolved & .2911 & .2180 & .1695 & .1385\\
			\hline
			HILL & .3449 & .2596 & .2063 & .1574\\
			HILL-evolved & .3559 & .2700 & .2241 & .1654\\
			\hline
			SUNIWARD & .3245 & .2169 & .1573 & .1139\\
			SUNIWARD-evolved & .3376 & .2309 & .1709 & .1245\\
			\hline\hline
		\end{tabular}
	\end{center}
\end{table}

\begin{figure}[!t]
	\centering
	\includegraphics[width=\linewidth]{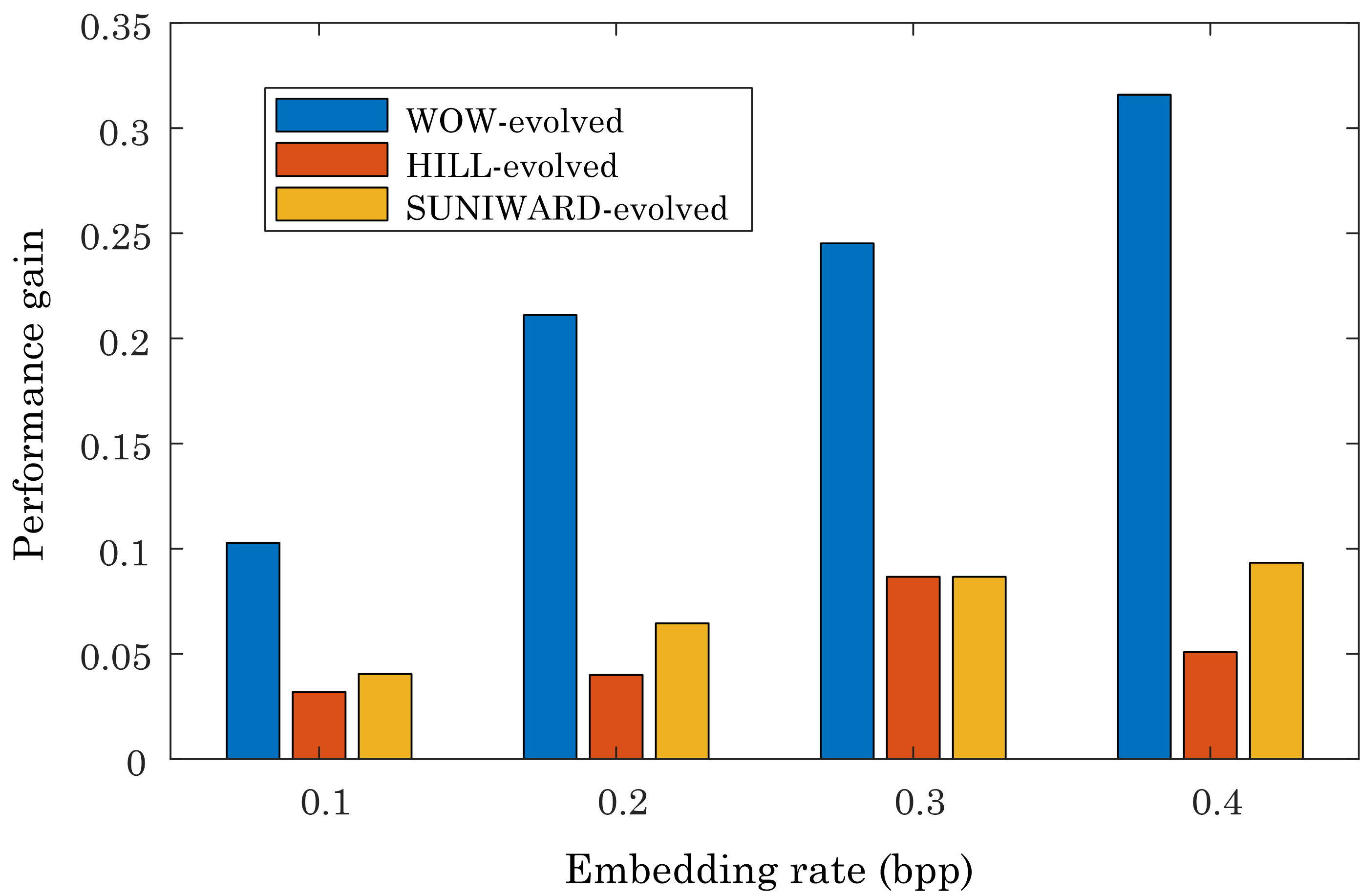}
	\caption{The relative performance gain for different evolved algorithms.}\label{gain}
\end{figure}

\subsection{Steganalysis Results}
The cost functions evolved from WOW, HILL, and SUNIWARD are referred to as the WOW-evolved, HILL-evolved, and SUNIWARD-evolved cost functions, respectively. Figure \ref{probMap} illustrates the embedding change probabilities of a cover image under a relative payload of 0.4 bpp using WOW, WOW-evolved, HILL, HILL-evolved, SUNIWARD and SUNIWARD-evolved cost functions. The probability map of modification is encoded in brightness, where brighter areas represent lower probabilities of embedding. The presented cover image contains numerous horizontal and vertical edges, along with textured regions. For clarity, a cropped and enlarged portion of the image is displayed.

A visual comparison between WOW and WOW-evolved reveals that the evolved version effectively suppresses grid-like artifacts introduced by WOW. Meanwhile, within complex textured areas of high embedding probability, the modification pattern exhibits a noise-like, stochastic appearance. HILL-evolved successfully mitigates the checkerboard pattern characteristic of HILL. It more precisely concentrates the embedding changes in texturally complex regions, where the modifications likewise blend in as a natural, noise-like texture. Finally, SUNIWARD-evolved demonstrates enhanced sensitivity to vertical details compared to SUNIWARD, assigning a higher embedding probability to those structural elements, demonstrating the potential of evolved function.

We report the performance of the evolved algorithms against steganalysis. The performance is evaluated across various embedding rates ranging from 0.1 bpp to 0.4 bpp with a step of 0.1. The detection error rate in Eq. (13) is adopted as the indicator. The experimental setting is the same as the evolutionary process, e.g., SRNet is used as the steganalyzer, and the ratio between training set, validation set and test set is 5:1:4. The experimental results shown in Table 1 indicate that all evolved algorithms consistently outperform their original counterparts, exhibiting enhanced security across every embedding strength and improved resistance to steganalysis. To quantify this improvement, we determine the (relative) performance gain, defined as $\Delta^\uparrow = (P_\text{E}^\star-P_\text{E}^\circ) / P_\text{E}^\circ$, where $P_\text{E}^\star$ and $P_\text{E}^\circ$ correspond to the evolved algorithm and raw version.

As shown in Figure \ref{gain}, despite minor fluctuations, the evolved algorithms exhibit a generally upward trend in terms of relative gain as the embedding rate increases, which aligns with the fact that steganographic security typically degrades at higher embedding rates, thereby leaving greater room for enhancement through evolutionary optimization. Overall, the results validate the generalizability of the proposed evolutionary framework across diverse steganographic strategies and highlight the ability of LLMs to automatically design steganographic schemes with superior security.

\subsection{Efficiency}
\begin{figure}[!t]
	\centering
	\includegraphics[width=\linewidth]{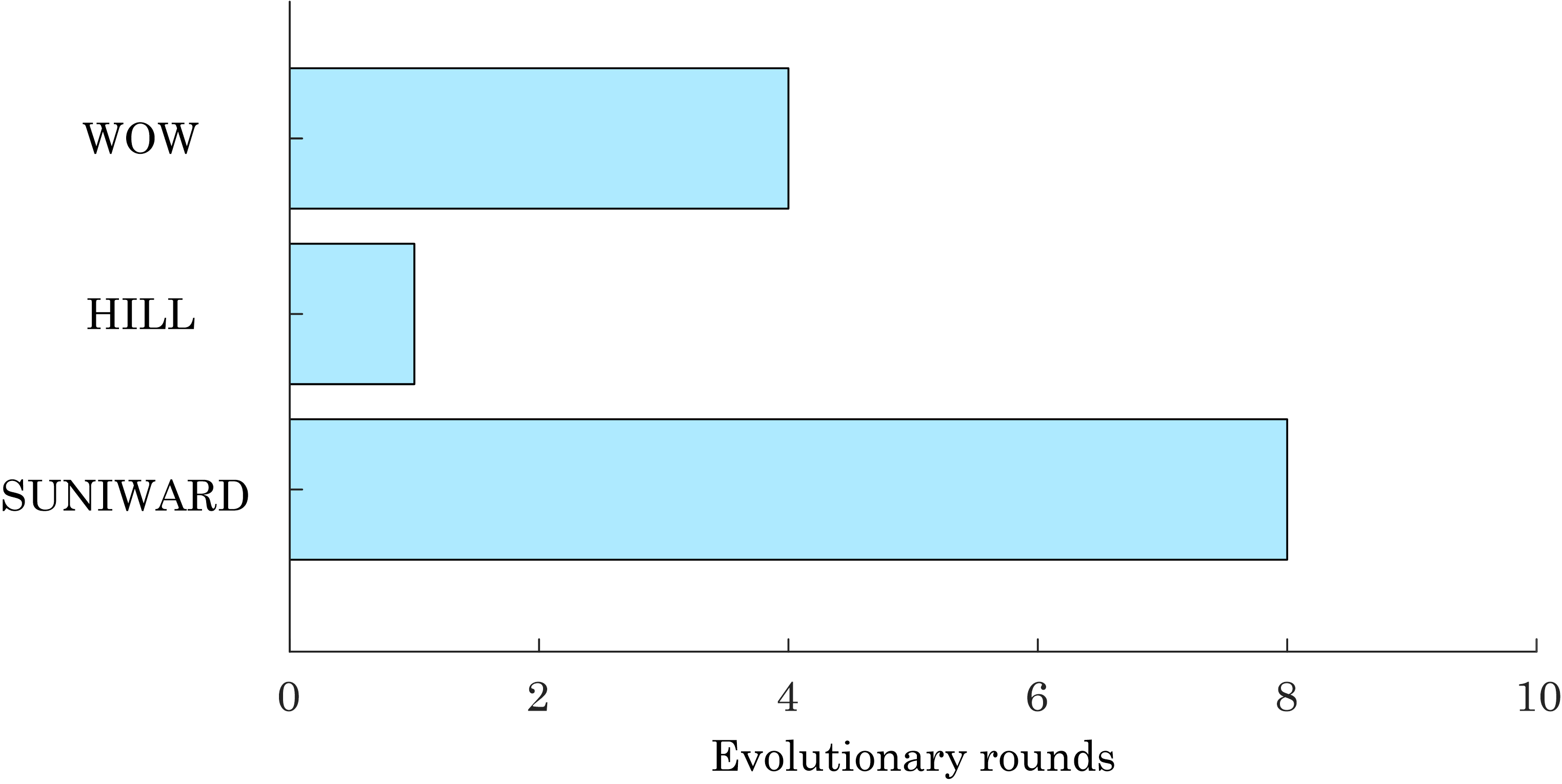}
	\caption{The number of evolutionary rounds required for each algorithm to obtain the final evolved cost function.}\label{rounds}
\end{figure}

We further investigate the efficiency of the proposed framework in generating cost functions, which directly influences computational cost and deployment feasibility. As shown in Figure \ref{rounds}, all evolved cost functions require a small number of evolutionary rounds. The results indicate that the proposed framework can produce high-performance functions without requiring extensive optimization, thereby significantly reducing computational overhead and underscoring its practical utility in applications.

\section{Conclusion and Future Work}
We introduce a novel reasoning-based framework that leverages LLMs to design cost functions of steganography. The proposed two-stage iterative framework first employs LLMs to evolutionarily generate and diversify candidate functions from structured prompts. The subsequent stage then conducts a more rigorous and accurate assessment to identify the superior cost functions while also providing feedback to the generation stage for further refinement. In contrast to experience-based methods that depend heavily on expert knowledge and learning-based approaches that require extensive training data yet lack interpretability, our method constructs novel cost functions in the form of executable programs by leveraging the reasoning capabilities of LLMs.

Experimental results demonstrate that the evolved cost functions consistently outperform their original versions across multiple embedding rates in resisting steganalysis. The evolution process is remarkably efficient, requiring only minimal rounds to produce evolved functions that effectively reduce artifacts and concentrate modifications more intelligently in complex textures. Although the cost functions generated in our experiments have not yet surpassed all existing benchmarks, our LLM-based framework demonstrates strong potential for automating the design of competitive steganographic algorithms. Future work will include augmenting LLM with a curated knowledge base of steganographic principles, as well as optimizing both LLM prompting strategies and evolutionary search mechanisms to fully realize this potential.

This work enables AI to design cost function through human-understandable text generation, which already opens up possibilities for the future, where AI fully replaces humans in the design of steganography algorithms. Furthermore, this framework is generalizable and can be extended to address other forensic problems.

\section{Acknowledgment} 
This study was financially supported by the Nanning ``Yong Jiang'' Program under Grant Number RC20250102, Science and Technology Commission of Shanghai Municipality under Grant Number 24ZR1424000, and Xizang Autonomous Region Central Guided Local Science and Technology Development Fund Project under Grant Number XZ202401YD0015. 



\onecolumn

\appendix



\section*{Supplementary material}

## Appendix A: Schematic diagram of the structured code for the steganographic workflow

```python
import numpy as np
# Load the essential libraries needed for computations and embedding operations.
......

def EMBED(cover, embedding_rate):
# This function embeds secret data into the cover image by modifying pixels according to
# their calculated change probabilities, ensuring the desired embedding rate is achieved
# and producing the corresponding stego image.
    ......
    return stego

@funsearch.evolve
def compute_cost_adjusted(cover, wet_cost):
# This function generates pixel-wise distortion cost matrices for a given cover image in
# the spatial domain. Guided by a large wet cost parameter, it assigns high costs to pixels
# that should not be modified and computes the cost of changing each pixel by +1 or -1
# under a ternary embedding scheme. These costs guide the embedding process to minimize
# distortion while avoiding sensitive regions.
    ......
    return rho_p1, rho_m1

def probability(rho_p1, rho_m1, embedding_rate, n):
# This function performs an iterative search using both exponential growth and binary
# search strategies to determine the optimal Lagrange multiplier λ that achieves the
# desired embedding rate given the cost matrices. It returns the final pixel modification
# probabilities.
    ......
    return (p_p1, p_m1), lbda

def average_payload(lbda, rho_p1, rho_m1, p_p1, p_m1):
# This function calculates the expected message payload and entropy based on either cost
# values and a Lagrange multiplier λ, or directly from given embedding probabilities.
    ......
    return (p_p1, p_m1), entropy

@funsearch.run
def evaluate(instances):
# This function executes the full embedding and testing pipeline. It embeds secret data
# into cover images using the EMBED function, saves the resulting stego images, and then
# evaluates their detectability using pre-trained steganalysis models. The function reports
# average classification accuracy and computes the final test score.
    ......
    return score
```

The framework accepts structured code depicting the entire workflow to comprehend the evolutionary search's target and procedure. The use of decorators enables the framework to automatically identify the two key functions. The `compute_cost_adjusted` function, which calculates the modification cost for each pixel in a cover image, is the target of the evolutionary optimization. The `evaluate` function serves as the main driver, performing embedding and subsequent evaluation with pre-trained steganalyzers by integrating the functions together. During use, the green ellipsis will be replaced with the specific steganography implementation.

## Appendix B: Python version of WOW-evolved cost function

```python
def compute_cost_adjusted(cover_spatial: np.ndarray, wet_cost: float=10000000000.0)
        -> typing.Tuple[np.ndarray]:
    d0 = (1 / 2) * np.array([1, 1])
    d1 = (1 / 2) * np.array([1, -1])
    G = [
        np.outer(d1, d0),
        np.outer(d0, d1),
        np.outer(d1, d1),
        np.outer(d0, d0),
        2 * np.eye(2)
    ]
    pad_size = max([g.shape[0] for g in G])
    cover_extended = np.pad(cover_spatial, pad_width=(pad_size, pad_size), mode='reflect')
    influence_maps = []
    for gIndex, g_matrix in enumerate(G):
        convolved = scipy.ndimage.convolve(cover_extended, g_matrix, mode='nearest')
        filtered_influence = gaussian_filter(np.abs(convolved), sigma=0.8)
        shift = g_matrix.shape[0] // 2
        rolled_influence = np.roll(np.roll(filtered_influence, shift, axis=0), shift, axis=1)
        influence_maps.append(rolled_influence[pad_size:-pad_size, pad_size:-pad_size])
    csf_weights = np.array([1.8, 1.4, 1.6, 1.0, 0.9])
    p = -2.5
    accumulated_influence = np.sum([csf_weights[i] * influence_maps[i] ** p for i in
                                    range(len(G))], axis=0)
    rho = np.where(accumulated_influence != 0, np.power(accumulated_influence, -1/p),
                   wet_cost)
    rho[rho < 1e-12] = wet_cost
    rho_p1 = np.where(cover_spatial > 254, wet_cost, rho)
    rho_m1 = np.where(cover_spatial < 1, wet_cost, rho)
    dynamic_threshold_p1 = np.percentile(rho_p1[rho_p1 < wet_cost], 95)
    dynamic_threshold_m1 = np.percentile(rho_m1[rho_m1 < wet_cost], 95)
    for i in range(rho_p1.shape[0]):
        for j in range(rho_p1.shape[1]):
            if rho_p1[i, j] > dynamic_threshold_p1:
                rho_p1[i, j] = wet_cost
    for i in range(rho_m1.shape[0]):
        for j in range(rho_m1.shape[1]):
            if rho_m1[i, j] > dynamic_threshold_m1:
                rho_m1[i, j] = wet_cost
    return rho_p1, rho_m1
```

The WOW-evolved cost function generated by the LLM in the form of Python code. The modifications marked in blue bold represent the key innovations of the WOW-evolved function relative to the original WOW.

# Appendix C: Python version of HILL-evolved cost function

```python
def compute_cost_adjusted(cover_spatial: np.ndarray, wet_cost: float = 10000000000.0) \
        -> typing.Tuple[np.ndarray]:
    H_KB = np.array([
        [-1, 2, -1],
        [2, -4, 2],
        [-1, 2, -1]
    ], dtype='float32')
    padsize = max(H_KB.shape)
    cover_shape = cover_spatial.shape
    cover_padded = np.pad(cover_spatial, pad_width=padsize, mode='symmetric')
    cover_padded = np.asarray(cover_padded, dtype='float64')
    I1 = scipy.ndimage.convolve(cover_padded, H_KB, mode='reflect')
    L1 = np.ones((3, 3), dtype='float64') / 9
    I2 = scipy.ndimage.convolve(np.abs(I1), L1, mode='reflect')
    if H_KB.shape[0] % 2 == 0:
        I2 = np.roll(I2, 1, axis=0)
    if H_KB.shape[1] % 2 == 0:
        I2 = np.roll(I2, 1, axis=1)
    crop_row_start = (I2.shape[0] - cover_shape[0]) // 2
    crop_row_end = crop_row_start + cover_shape[0]
    crop_col_start = (I2.shape[1] - cover_shape[1]) // 2
    crop_col_end = crop_col_start + cover_shape[1]
    I2 = I2[crop_row_start:crop_row_end, crop_col_start:crop_col_end]
    rho = 1 / (I2 + 1e-10)
    smoothed_rho = gaussian_filter(rho, sigma=3)
    smoothed_rho[np.isinf(smoothed_rho) | np.isnan(smoothed_rho) |
                 (smoothed_rho > wet_cost)] = wet_cost
    rho_p1 = np.copy(smoothed_rho)
    rho_m1 = np.copy(smoothed_rho)
    for i in range(cover_shape[0]):
        for j in range(cover_shape[1]):
            if cover_spatial[i, j] >= 255:
                rho_p1[i, j] = wet_cost
            if cover_spatial[i, j] <= 0:
                rho_m1[i, j] = wet_cost
    return rho_p1, rho_m1
```

The HILL-evolved cost function generated by the LLM in the form of Python code. The modifications marked in blue bold represent the key innovations of the HILL-evolved function relative to the original HILL.

## Appendix D：Python version of SUNIWARD-evolved cost function

```python
def compute_cost_adjusted(cover_spatial: np.ndarray, wet_cost: float = 10000000000.0) \
        -> typing.Tuple[np.ndarray]:
    sgm = 1
    hpdf = np.array([
        -.0544158422, .3128715909, -.6756307363, .5853546837,
        .0158291053, -.2840155430, -.0004724846, .1287474266,
        .0173693010, -.0440882539, -.0139810279, .0087460940,
        .0048703530, -.0003917404, -.0006754494, -.0001174768])
    lpdf = (-1) ** np.arange(len(hpdf)) * np.flip(hpdf)
    F = []
    F.append(np.outer(lpdf, hpdf))
    F.append(np.outer(hpdf, lpdf))
    F.append(np.outer(hpdf, hpdf))
    sizes = np.array([f.shape for f in F])
    padSize = np.max(sizes, axis=0)
    cover_padded = np.pad(cover_spatial, pad_width=padSize, mode='symmetric')
    cover_padded = np.asarray(cover_padded, dtype='float64')
    cover_shape = cover_spatial.shape
    direction_weights = np.array([0.5, 1.0, 0.5])
    xi = [None] * 3
    for fIndex in range(3):
        R = scipy.ndimage.convolve(cover_padded, F[fIndex], mode='reflect')
        abs_filter = np.rot90(np.abs(F[fIndex]), 2)
        if fIndex == 1:
            adapt_filter = abs_filter * 1.5
        else:
            adapt_filter = abs_filter
        xi[fIndex] = scipy.ndimage.convolve(1 / (np.abs(R) + sgm), adapt_filter,
                                            mode='reflect')
        if F[fIndex].shape[0] % 2 == 0:
            xi[fIndex] = np.roll(xi[fIndex], shift=1, axis=0)
        if F[fIndex].shape[1] % 2 == 0:
            xi[fIndex] = np.roll(xi[fIndex], shift=1, axis=1)
        row_offset = (xi[fIndex].shape[0] - cover_shape[0]) // 2
        col_offset = (xi[fIndex].shape[1] - cover_shape[1]) // 2
        xi[fIndex] = xi[fIndex][row_offset:-row_offset, col_offset:-col_offset]
    rho = (xi[0] * direction_weights[0] + xi[1] * direction_weights[1] +
           xi[2] * direction_weights[2])
    rho[np.isinf(rho) | np.isnan(rho) | (rho > wet_cost)] = wet_cost
    rho_p1 = np.copy(rho)
    rho_m1 = np.copy(rho)
    threshold_variation = 5
    for i in range(cover_spatial.shape[0]):
        for j in range(cover_spatial.shape[1]):
            if cover_spatial[i, j] >= 255 - threshold_variation:
                rho_p1[i, j] = wet_cost
            if cover_spatial[i, j] <= 0 + threshold_variation:
                rho_m1[i, j] = wet_cost
    return rho_p1, rho_m1
```

The SUNIWARD-evolved cost function generated by the LLM in the form of Python code. The modifications marked in blue bold represent the key innovations of the SUNIWARD-evolved function relative to the original SUNIWARD.

 

\end{document}